\PassOptionsToPackage{usenames,dvipsnames}{xcolor}
\documentclass[conference]{IEEEtran}

\IEEEoverridecommandlockouts

\usepackage{cite}
\usepackage{amsmath,amssymb,amsfonts}
\usepackage{algorithmic}
\usepackage{graphicx}
\usepackage{subcaption}
\usepackage{textcomp}
\usepackage{url}
\usepackage[usenames,dvipsnames]{xcolor}
\usepackage{float}
\usepackage[normalem]{ulem}
\usepackage{booktabs}
\usepackage[strings]{underscore}
\usepackage{enumitem}
\usepackage[numbers]{natbib}
\usepackage[switch,columnwise]{lineno}
\usepackage{mathabx}
\usepackage{pifont}

\newcommand{\nassos}[1] { \textcolor{teal}{{\bf AM: }{#1}}}

\newcommand{\xmark}{\ding{55}}%

\def\BibTeX{{\rm B\kern-.05em{\sc i\kern-.025em b}\kern-.08em
    T\kern-.1667em\lower.7ex\hbox{E}\kern-.125emX}}

\begin{document}

\title{Towards Practical Fabrication Stage Attacks\\Using Interrupt-Resilient Hardware Trojans\\
{\footnotesize 2024 IEEE International Symposium on Hardware Oriented Security and Trust (HOST)}
\thanks{© 2024 IEEE. Personal use of this material is permitted. Permission from IEEE must be
obtained for all other uses, in any current or future media, including reprinting/republishing this material for advertising or promotional purposes, creating new collective works, for resale or redistribution to servers or lists, or reuse of any copyrighted component of this work in other works.}}

\author{\IEEEauthorblockN{Athanasios Moschos}
\IEEEauthorblockA{
\textit{Georgia Institute of Technology}\\
Atlanta, Georgia, USA \\
amoschos@gatech.edu}
\and
\IEEEauthorblockN{Fabian Monrose}
\IEEEauthorblockA{
\textit{Georgia Institute of Technology}\\
Atlanta, Georgia, USA \\
fabian@ece.gatech.edu}
\and
\IEEEauthorblockN{Angelos D. Keromytis}
\IEEEauthorblockA{
\textit{Georgia Institute of Technology}\\
Atlanta, Georgia, USA \\
angelos@gatech.edu}
}

\maketitle
\thispagestyle{plain}
\pagestyle{plain}

\begin{abstract}
We introduce a new class of hardware trojans called \textit{interrupt-resilient trojans} (IRTs).
Our work is motivated by the observation that hardware trojan attacks on CPUs, even under favorable attack scenarios (\textit{e.g.,} an attacker with local system access), are affected by unpredictability due to non-deterministic context switching events.
As we confirm experimentally, these events can lead to race conditions between trigger signals and the CPU events targeted by the trojan payloads (\textit{e.g.,} a CPU memory access), thus affecting the reliability of the attacks.
Our work shows that interrupt-resilient trojans can successfully address the problem of non-deterministic triggering in CPUs, thereby providing high reliability guarantees in the implementation of sophisticated hardware trojan attacks.
Specifically, we successfully utilize IRTs in different attack scenarios against a Linux-capable CPU design and showcase its resilience against context-switching events.
More importantly, we show that our design allows for seamless integration during fabrication stage attacks.
We evaluate different strategies for the implementation of our attacks on a tape-out ready high-speed RISC-V microarchitecture in a 28nm commercial technology process and successfully implement them with an average overhead delay of only 20 picoseconds, while leaving the sign-off characteristics of the layout intact.
In doing so, we challenge the common wisdom regarding the low flexibility of late supply chain stages (\textit{e.g.,} fabrication) for the insertion of powerful trojans.
To promote further research on microprocessor trojans, we open-source our designs and provide the accompanying supporting software logic.
  

\end{abstract}


\section{Introduction}
\label{sec:introduction}

Hardware trojans (HTs)~\cite{KRRJRKTM10} have become a topic of increased attention, due to the covertness of their nature and their potential for malicious exploitation of globalized supply chains.
Motivated adversaries leverage the opacity and the geopolitics of modern semiconductor supply chains to target consumer or military grade electronics. 
In their most clandestine form, hardware trojans are forged by intentional modifications made directly to the physical layout of integrated circuits (ICs).
Their presumed out in the wild occurrences~\cite{IEEEX:ADE08, DBLP:SkorobogatovW12}, although rare, have sparked great interest in the defensive hardware security domain, with different detection strategies spanning from logic testing~\cite{CRSWFPSPCBW09, IEEE:LYMP21TAR} to detection through side-channels~\cite{ACM:LYMP21MAX,ACM:HYBSMP16,DBLP:NguyenCPZ19} and scanning electron microscopes~\cite{EPTMSBCKAMCP23, VNLHSQRMTSHWDLANTM18, CFLMPFJJATA15}.
The shared objective  is to enhance the security of existing supply chain stages and make them more resilient to attackers looking to insert {small, but meaningful, modifications} in host systems of interest.  

At the same time, research on the offensive side~\cite{DBLP:KTCGJZ08, DBLP:TM14, DBLP:YHDAS16, DBLP:KAFP19, De2020HarTBleedUH, DBLP:HS21, AMVKADK, ACM:VGHGSPJR, ACM:AHTPSPGS, KDJC23} has shed light on the various mechanisms attackers can use at different stages of the chips' design life cycle. Indeed, the large body of offensive research  
~\cite{DBLP:KAFP19, De2020HarTBleedUH, DBLP:KTCGJZ08, KDJC23, DBLP:TM14, DBLP:YHDAS16} has advanced our collective understanding of different capabilities of varied hardware trojan designs,
culminating in new lines of inquiry in the practical implementation of attacks~\cite{DBLP:YHDAS16, De2020HarTBleedUH, KDJC23} or the development of frameworks~\cite{ACM:VGHGSPJR} that are more and more powerful.

\par To date, much of that research~\cite{Xue2020TenYO} has focused on the stealthiness aspect of trojan attacks, where 
the number of trojan gates and nets partaking in the malicious modification is the key metric used to characterize a trojan's stealthiness~\cite{DBLP:TM14, DBLP:YHDAS16, DBLP:HS21, DBLP:KAFP19}.
Existing literature on hardware trojan attacks against CPUs~\cite{DBLP:KTCGJZ08, De2020HarTBleedUH, ACM:VGHGSPJR, KDJC23} often assumes that attackers are able to excite trojans hidden in the microarchitecture through the execution of malicious code on the same system.
Under this threat model, the malicious binary can surreptitiously trigger the trojan to activate the payload and deliver the desired outcome.
Aside from inducing denial of service effects~\cite{ACM:SPMGFSTRD, AMVKADK, IEEEX:ADE08}, more intricate HT payloads could modify some aspect of the CPU state space to the attacker's benefit, such as temporarily disabling memory access control~\cite{DBLP:KTCGJZ08, DBLP:TM14} or making type-safe software vulnerable~\cite{KDJC23}.

\par However, an often overlooked factor is how the interplay between software and hardware in modern microprocessor systems can pose challenges for the correct execution of a hardware trojan attack. 
Specifically, modern CPUs are complex finite state machines that handle numerous unpredictable, asynchronous events (\textit{e.g.,} interrupts). 
Thus, it is often the case that between the triggering of the trojan and the delivery of the effect, the CPU state can change, and if that change is not properly accommodated for, it can lead to undesirable side effects. 
The unintended consequences (\textit{e.g.,} a crash) cause unexplainable system behaviors that raise attention and inevitably lead to further scrutiny of the target device.
\textit{We therefore consider attack reliability to be as equally as important as a trojan's size when it comes to stealthiness.}


Despite the abundance of research on CPU-based hardware trojan designs~\cite{DBLP:TM14, DBLP:YHDAS16, DBLP:KAFP19, DBLP:HS21}, there has been a lack of research on how the subtleties of asynchronous events can undermine the stealthiness of attacks.
Motivated by this observation, we shed light on the wide variety of context switching (CS) events that can occur during the execution of CPU HT attacks.
Moreover, from the stand-point of a fabrication stage attacker, we examine two CS-resilient hardware trojan designs that can be efficiently implemented in tape-out ready layouts.
We make the following contributions:
\begin{itemize}
    
    \item We introduce a new class of hardware trojans, called \textit{interrupt-resilient trojans} (IRTs), that tackles context-switching events and manages the reliable delivery of the time-critical trigger signal.
    
    \item We show how IRTs can be used to perform successful attacks against the availability and integrity of a CPU design under different context switching scenarios. 

    \item To showcase the diversity of options available to the attacker, we identify parts of the CPUs' microarchitecture that can be used to insert IRTs at the fabrication stage.
    
    \item We emulate a fabrication stage attacker and insert IRTs in a tape-out ready CPU layout.
    
\end{itemize}

Overall, our work aims to bring attention to practical challenges in designing hardware trojan attacks and to revisit longstanding assumptions about the presumed lack of flexibility when implementing fabrication stage hardware trojans~\cite{Jacob2014HardwareTC}.

\paragraph*{Paper Organization}
Section~\ref{sec:background} provides an overview of contemporary CPU HT attacks, including their operational threat models, as well as a primer on how context switching has been viewed in the HT literature.
Section~\ref{sec:threat_model} outlines the threat model we consider for the implementation of our HT attacks. 
Section~\ref{sec:our_approach} introduces interrupt-resilient trojans with their design characteristics (Section~\ref{sec:irt_trojans}) and operational models (Section~\ref{sec:irt_operational}), as well as practical considerations when realizing them in silicon during fabrication stage attacks (Section~\ref{sec:tt_methodologies}).
In Section~\ref{sec:eval_cs} we evaluate IRTs in the execution of attacks against a modern RISC-V microarchitecture under diverse context switching events.
We evaluate the impact of IRTs' design on a tape-out ready layout of a RISC-V microarchitecture in Section~\ref{sec:eval_asic}.
Section~\ref{sec:mitigations} discusses possible mitigations against IRTs.
We conclude in section~\ref{sec:conclusions}.

\section{Background and Related Work}
\label{sec:background}
        


\label{sec:htrojans}
    \par Modern day CPUs \cite{DaBo20} are sophisticated general-purpose hardware that can seamlessly accommodate a wide variety of operating systems (OSes) and software.
The executed software heavily relies on the proper implementation of the underlying hardware to produce correct results, but it has limited means of tracing erroneous realizations of the hardware itself. As such, the flexibility afforded by modern CPUs comes at price --- that is, the lack of an explicit "contract" between the software and the underlying hardware makes the latter a prime target for the inclusion of hardware trojans.

\par In computer security parlance, an HT consists of a trigger circuit and a payload circuit.
The \textit{trigger circuit} is tasked with monitoring for predefined conditions and initiating the payload delivery when these conditions are met.
In general, trigger circuits must be designed in a way that they remain hidden during the testing phase or normal chip operation, but are also easy to excite at the attacker's behest.
A trigger circuit meeting these requirements is considered to be highly reliable.
The second vital component is the \textit{payload circuit}, which is responsible for actuating an attack on the host system through the modification of the system's native capabilities.

\par However a trojan is not an organic part of the CPU's design, unless it is added during the initial specification or the physical implementation of the microarchitecture.
Only stages prior to fabrication~\cite{Jacob2014HardwareTC} provide flexibility with respect to smoothly integrating microarchitectural modifications, whether that is through high-level specification adjustments or low-level modifications of the RTL logic at the more lenient front-end and back-end stages.  

\par Irrespective of the insertion stage, trojan payloads need to interfere with native CPU events.
A malicious process, referred from now on as {\it handling process}, is also natively executed on the CPU in order to exercise the HT's capabilities. 
Therefore, it is important that during the design of the trojan, the adversary considers ways that allow for effective interaction with the handling software, so that the HT can accurately target CPU events of interest.
If this collaboration cycle breaks down, the trojan runs the risk of being discovered.
As such, covert collaboration between the handling software and the HT's functionality is critically important for reliably delivering an attack that flies under the radar.





\label{sec:related_work}
\begin{table}[H]
\footnotesize
\centering
\scalebox{0.9}{
\begin{tabular}{c c c c c}
    \toprule
    \textbf{Publication}& \textbf{\shortstack{Insertion\\Stage}}&\textbf{\shortstack{Access\\Vector}}& \textbf{\shortstack{Triggering\\Medium}}& \textbf{\shortstack{Addressed\\Context\\ Switching}}\\
    \midrule
    \cite{DBLP:KTCGJZ08}& design& mix& mix& not applicable\\
    \cite{DBLP:TM14}& fabrication& local& process& partially\\
    \cite{DBLP:YHDAS16}& fabrication& local& process& \textcolor{Red}{\xmark}\\
    \cite{DBLP:KAFP19}& mix& local& process& \textcolor{Red}{\xmark}\\
    \cite{De2020HarTBleedUH}& fabrication& local& process& \textcolor{Red}{\xmark}\\
    \cite{DBLP:HS21}& design& -& -& not applicable\\
    \cite{AMVKADK}& fabrication& network& packet& \textcolor{Red}{\xmark}\\
    \cite{ACM:VGHGSPJR}& fabrication& local& process& \textcolor{Red}{\xmark}\\
    \cite{ACM:AHTPSPGS}& fabrication& -& -& \textcolor{Red}{\xmark}\\
    \cite{ACM:SPMGFSTRD}& fabrication& -& -& \textcolor{Red}{\xmark}\\
    \cite{KDJC23}& design& mix& process& not applicable\\
    \hline
    Our Work& fabrication& local& process& \textcolor{ForestGreen}{\checkmark}\\
    \bottomrule
\end{tabular}
}\\
\scriptsize{Insertion Stage: design, fabrication, mix\\
Triggering Method: network packet, user process, mix\\
Access Vector: network, local, mix}\\

\caption{Contemporary CPU Hardware Trojan Attacks}
\label{table:literature}
\end{table}

\paragraph*{Contemporary Attacks}
State of the art research on CPU HT attacks~\cite{DBLP:KTCGJZ08, DBLP:TM14, DBLP:YHDAS16, DBLP:KAFP19, De2020HarTBleedUH, DBLP:HS21, AMVKADK, ACM:VGHGSPJR, ACM:AHTPSPGS, ACM:SPMGFSTRD, KDJC23} span different insertion stages (e.g., design vs fabrication) and access level (local vs remote). Table~\ref{table:literature} summarizes the most germane work in this area. In these works, the size of the HT implementations --- namely, the number of trojan gates and signals that are necessary to implement the trojan --- is  used to justify arguments about stealthiness. Intuitively, 
a smaller size translates to a stealthier HT with increased probabilities of remaining undetected during testing phase or after the host chip's deployment. 
Unfortunately, the operational aspect of a HT implementation is usually considered orthogonal to its stealthines, and therefore out of the scope of works like~\cite{DBLP:HS21, ACM:AHTPSPGS, ACM:SPMGFSTRD}, which lack a discussion about the conditions under which their HTs can operate. 
For the works that do (\textit{e.g.,}~\cite{DBLP:YHDAS16, DBLP:KAFP19, De2020HarTBleedUH, AMVKADK, ACM:VGHGSPJR, KDJC23}), they usually lack an in-depth discussion of how the intricacies of the CPU's operational reality can undermine the effectiveness of their proposed attacks.
Indeed, several approaches simply assume that attackers are readily able to execute malicious code on the target {\it without any interruption}.


\label{sec:ht_execution}
    \paragraph*{Challenges in Executing a Hardware Trojan Attack}
Modern OSes segregate virtual memory into user and kernel space in order to provide memory and hardware protection from malicious or erroneous software behavior. The privileged OS kernel, as well as some device drivers, are strictly executed in a reserved memory area called the kernel space while software applications are executed in user space. Furthermore, to improve performance, modern microprocessors interface a variety of peripheral modules and allow for OS multitasking.  
Multitasking is accomplished through the operation of context switching, where the state of a process or thread is saved on interrupt and then later restored once the execution of the process is resumed~\cite{CSdefinition}.
Context switching can happen for a wide variety of reasons, including (but not limited to) multitasking, interrupt handling, user and kernel mode switching.
Undoubtedly, the assumption that attacks will execute without interruption is \textit{unrealistic}.

\par Research works that consider design stage attackers~\cite{DBLP:KTCGJZ08, DBLP:HS21, KDJC23} in Table~\ref{table:literature} allow for a smooth HT integration with the microarchitecture and avoid the challenges stemming from context switching. 
Unfortunately, among all the fabrication stage approaches, only the work by~\citet{De2020HarTBleedUH} has acknowledged the challenges posed by context switching for the implementation of an attack.
The work by~\citet{DBLP:TM14} is the only implementation that considers how context switching events could be used to bolster the HT's operation. 
Specifically, they show how CS events are leveraged by the injected HT, so that a malicious user process can gain access to an unauthorized memory address space.
Unfortunately, their implementation considers a specific attack and its correct execution is not guaranteed. 

\par By now it should be clear that regardless of which of the scenarios in Table~\ref{table:literature} is assumed, with respect to the attacker's access vector on the system, the handling software must deal with unpredictable interruptions from context switching.
It is precisely this uncertainty that we aim to tackle head on. Specifically, we argue that \textit{any realistic hardware trojan attack must accommodate for non-deterministic context switching events}, generated by the continuous interleaving of interrupts and applications in the execution chain of a microprocessor.

\section{Threat Model}
\label{sec:threat_model}
    \par Generally speaking, a hardware trojan may be implemented either at the front-end, back-end or fabrication phase of a chip's design cycle.
Within the security community, fabrication stage attacks are assumed to be more restrictive~\cite{Xue2020TenYO, De2020HarTBleedUH, DBLP:YHDAS16, DBLP:TM14, Jacob2014HardwareTC} in terms of the power afforded to an attacker (\textit{e.g.,} limited information about the underlying design, less flexibility about sign-off layout modifications). 
We question this longstanding assumption, and instead show that sophisticated HTs can indeed be implemented efficiently at this phase and lead to attacks equally as powerful as those enabled by insertion of HTs in design phases prior to fabrication~\cite{Jacob2014HardwareTC}.

\par Specifically, we adopt the threat model of a fabrication-stage attack, where a malicious entity inside a foundry gains access to the chip's GDSII file to introduce an IRT.
The GDSII format is the industry standard used in EDA to represent the planar geometric shapes and relevant information of the laid-out and routed IC in a binary database file format.
This file is used as input in the fabrication machinery to generate the photomasks used in the photolithography process.
We assume that the design process up to the generation of the GDSII by the design house is completely trusted and any malicious alterations take place inside the foundry.
We assume the attacker in the fabrication facility has access to modern Electroninc Design Automation (EDA) tools, as well as the standard cell libraries, the process design kit (PDK) and any hard macros (\textit{e.g.,} SRAMs) used for the generation of the victim layout.

\par Since the chip information represented in the GDSII is at a sign-off level (that is, there can be no timing or manufacturing violations), an attacker must  be extremely careful with the layout modifications required for a hardware trojan implementation.  
At this phase, we assume the attacker has the ability to add attack-circuits to open spaces of the laid-out design, for example, places occupied by filler cells for DFM (design for manufacturing) purposes.
This is a widely accepted practice in the hardware security community with regards to fabrication stage attacks~\cite{De2020HarTBleedUH, DBLP:YHDAS16, ACM:AHTPSPGS, AMVKADK, ACM:VGHGSPJR, DBLP:TM14}.

\par We also remind the reader that it is  typical practice for design houses to provide abstract information to foundries about the functionality of the chip delivered for fabrication.
Therefore, a malicious entity inside the foundry can utilize this shared information, as well as public domain data with respect to the design houses' clients and deduce the Instruction Set Architecture (ISA) implemented by the chip. Moreover, it is reasonable to assume that an attacker can  extract the gate-level netlist~\cite{RRSLYJPD2020, DBLP:YHDAS16} of the targeted layout and use it to search for suitable victim flip-flops or routed signals by running a variety of test-benches in an HDL simulator (\cite{DBLP:YHDAS16}) or by applying reverse engineering methods to reconstruct the high-level functionality of the underlying modules~\cite{MTZSJY2016}.
Taken as a whole, we show that an adversary familiar with fabrication processes, and experienced in both analog and digital IC design and layout, can successfully insert an IRT.

\par With respect to the trojan operation, we consider an attacker 
that is able to interface with the modified CPU after its on field deployment inside a host system and deploy handling software in a fashion similar to the local access scenarios considered in Section~\ref{sec:ht_execution} (\textit{i.e.,} having user access privileges in an infected VM host). 
From then on, the attacker can exercise the HT's capabilities, leveraging the reliability offered by the interrupt resilient triggering mechanism.

\section{Our Approach}
\label{sec:our_approach}
    \begin{figure*}[ht]
    \centering
    \includegraphics[scale=0.25]{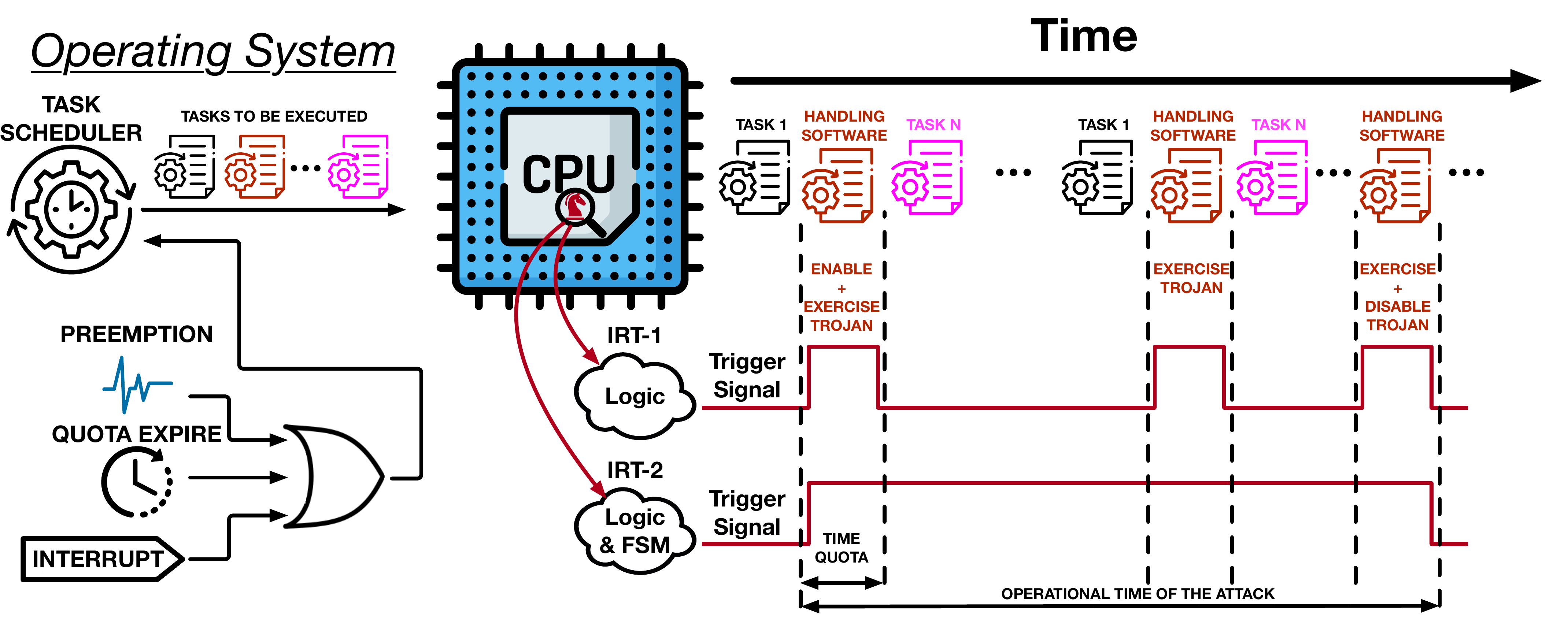}
    \caption{\centering Time slicing of the CPU's execution time during the execution of a hardware trojan attack.}
    \label{fig:scheduling}
    \vspace{-1em}
\end{figure*}

\par The goal of a HT attack against a CPU is to reliably deliver the payload, and attack success therefore should be unaffected by the normal operation of the microprocessor.
For illustrative purposes, consider the scenario proposed by~\cite{DBLP:YHDAS16}  where an attacker wishes to perform a privilege escalation attack and elevate the rights of a malicious process.
To achieve that, a payload must flip the state of the privilege bits of the microarchitecture to change the execution state of the handling process.
In this scenario, the operational time of the handling process is mostly divided between the time needed to execute instructions that activate the capacitive based trigger and the time spent executing instructions that take advantage of the HT's payload.
In practice, the operational period of the handling process is not contiguous but rather a chain of time slices because of interleaved multitasking and kernel switching phenomena, as seen in Figure~\ref{fig:scheduling}.
Consequently the noise induced in the operational time of the attack is significant and not easily predicted.
Therefore upon context switching back to the handling process, both the state of the privilege bits (payload), as well as the code location from which CPU will resume execution, are unknown.
If the handling process resumes execution and the trigger circuit fails to enable the payload on time (or keep it enabled), the attack will fail and this might jeopardize the HT's stealthiness.
\textit{For these reasons, we view the trigger signal as the most time critical aspect of hardware trojan designs.}

\par Thus, we first focus our attention on how to reliably deliver a trigger signal under diverse context switching events.
Afterwards, we show how our proposed trigger mechanisms can be combined with payloads, during the fabrication stage, to produce attack vectors equally powerful to those considered in design stage attacks (\textit{e.g.,}~\cite{KDJC23}).

\begin{table*}
\centering
\captionsetup{justification=centering}
\subcaptionbox{\scriptsize{Suitable CPU host modules for IRT-1 and IRT-2.}}{
\begin{tabular}[b]{c | c c}
    \toprule
    \textbf{\shortstack{CPU\\ Modules}}&\textbf{\shortstack{Attacker\\Influenced}}&\textbf{\shortstack{Restoring\\ Architectural State}}\\
    \midrule
    Adders $\star$&\textcolor{ForestGreen}{\checkmark}& \textcolor{Red}{\xmark}\\
    Dividers&\textcolor{ForestGreen}{\checkmark}& \textcolor{Red}{\xmark}\\
    Multipliers&\textcolor{ForestGreen}{\checkmark}& \textcolor{Red}{\xmark}\\
    Program Counter&\textcolor{ForestGreen}{\checkmark}&\textcolor{ForestGreen}{\checkmark}\\
    Floating Point Registers&\textcolor{ForestGreen}{\checkmark}&\textcolor{ForestGreen}{\checkmark}\\
    General Purpose Registers $\star$&\textcolor{ForestGreen}{\checkmark}&\textcolor{ForestGreen}{\checkmark}\\
    \bottomrule
\end{tabular}
\label{table:cpu_attacker_infl}
}
\quad
\captionsetup{justification=centering}
\subcaptionbox{\scriptsize{CPU events lasting multiple clock-cycles.}}{
\begin{tabular}[b]{c}
    \toprule
    \textbf{\shortstack{Time Expensive Events\\(Microarchitecture Dependent)}}\\
    \midrule
    Number Division\\
    Page Table Walking $\star$\\
    Branch Missprediction\\
    Missaligned Memory Accesses\\
    Translation Lookaside Buffer Flush\\
    \bottomrule
\end{tabular}
\label{table:cpu_time_expensive}
}
\captionsetup{justification=centering}
\caption{Modules and CPU events in support of IRTs. \\The $\star$ symbol denotes the examples we used for our evaluation in Section~\protect\ref{sec:eval_main}.}
\label{table:cpu_tables}
\end{table*}

\subsection{Trigger Circuit Designs}
\label{sec:irt_trojans}
    \par To directly confront the aforementioned challenges, we introduce \textit{interrupt-resilient trojans} or IRTs.
Interrupt-resilient trojans overcome the non-deterministic CS events happening during the CPU's normal operation through unique CS-aware trigger designs.
In what follows, we characterize IRTs by their trigger mechanism and describe the rationales behind our design choices.
First we discuss a \textit{selectively-ready} trigger mechanism that delivers the trojan trigger only when instructions of the handling software are executed inside the CPU's pipeline.
The second solution features an \textit{always-ready} trigger mechanism that remains in a ready state throughout the attack, irrespective of the instructions executed inside the CPU's pipeline.
Both of the trigger mechanisms provide to HTs the right conditions to deliver the payloads in a reliable and timely manner during the execution of an attack. 

\subsubsection{Selectively-Ready Trigger (IRT-1)}
\label{sec:irt_solution1}
    \par As discussed earlier, in every context switch event the operating system is responsible for saving the state of the currently running process and restoring the state of the next process to be executed.
This non-stop cycle of state saves and restores is fundamental for the support of the context switching mechanism.
We leverage the effects of this OS procedure in the CPU's microarchitecture to create IRT-1, a trigger mechanism implemented inside modules that experience those effects.
The requirements for the hardware logic of an IRT-1 host module is that it should ({\it i}) participate in the restore procedure when the CPU switches back to the handling process, and ({\it ii}) be deliberately influenced through non-privileged instructions executed by the handling software for the activation and de-activation of the attack.
We exemplify those requirement in Section~\ref{sec:irt_operational}.

\subsubsection{Always-Ready Trigger (IRT-2)}
\label{sec:irt_solution2}
    \par For the second type of trigger mechanism, we follow an ``always-ready" strategy to overcome the adversities stemming from constant CS events.
Unlike other ``always-ON" HTs that have no trigger conditions~\cite{Xue2020TenYO} and are constantly operating, our ``always-ready" solution refers to the ability of the IRT-2 trigger mechanism to remain in a ready state {\it only} during the operational time of the attack.
The only prerequisite for a prospective IRT-2 host module is to contain hardware logic that can be readily influenced through non-privileged instructions executed by the handling software, similarly to IRT-1.
Next, we explain the structural and operational differences between the two.

\subsection{Structural \& Operational Variations}
\label{sec:irt_operational}
\par
\begin{figure}[h]
    \centering
    \includegraphics[scale=0.28]{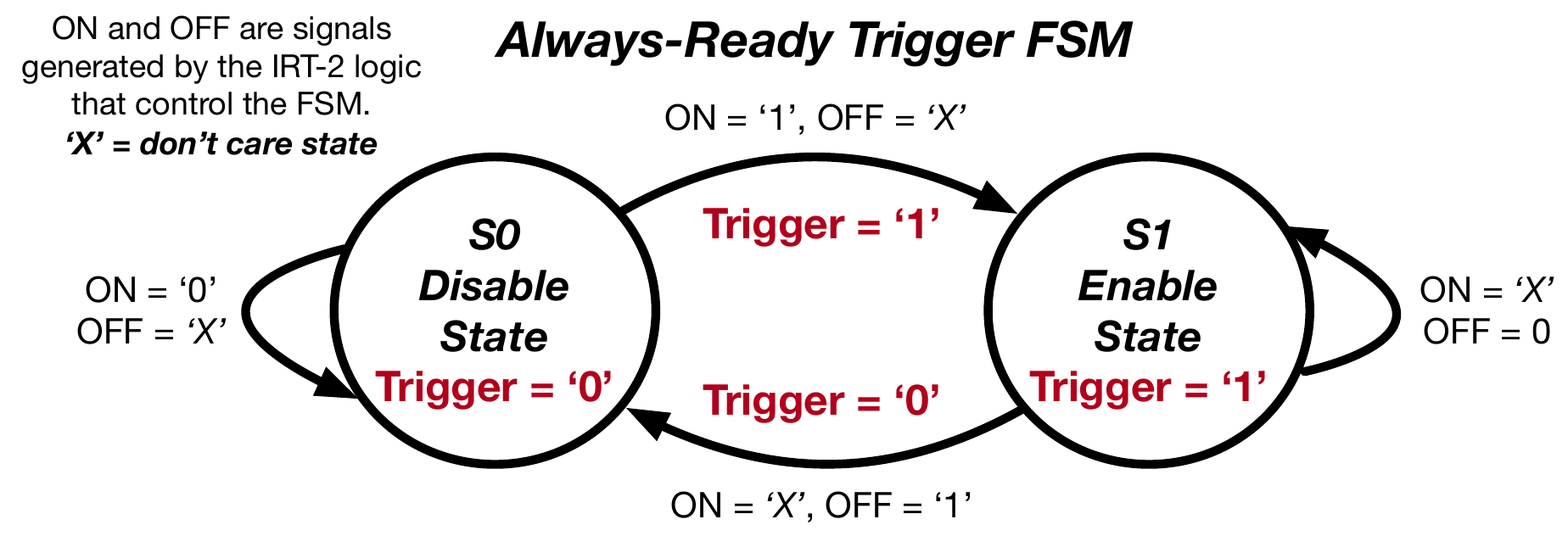}
    \caption{\centering Finite state machine of the ``Always-Ready" trigger.}
    \label{fig:always_readyFSM}
\end{figure}

 The attacker is able to select between our two interrupt-resilient trigger mechanisms depending on the HT design and the host system considerations.
A fundamental difference in the requirements of the two IRT mechanisms is that the hardware logic for host modules of IRT-1 need to support context switching procedures, while that is not a requirement for IRT-2.
In that sense, IRT-2 provides greater flexibility as it can ``infect" a bigger set of modules prevalent in a microarchitecture.
This is evident by comparing the two right columns in Table~\ref{table:cpu_tables}(a).
Nevertheless, the greater flexibility offered in the latter design comes at the cost of the extra support logic necessary to hold the trigger in an ON state.
This support logic is a simple FSM that can be seen in Figure~\ref{fig:always_readyFSM} and is responsible for the generation of the \textit{always-ready} trigger.

\par Naturally, this leads to a difference in the activation period between the two trigger solutions.
As illustrated in Figure~\ref{fig:scheduling} the IRT-1 solution delivers the trigger signal to the payload as long as the handling process is executed inside the CPU pipeline.
If the CPU has context switched to a different process, the trigger is automatically disabled until there is a context switch back to the handling software.
The handling process finally de-activates completely IRT-1 before ending.
On the other hand, the IRT-2 solution once activated stays enabled throughout the complete HT attack until the handling process terminates.
Therefore, the trigger signal of IRT-2 is constantly delivered to the payload, irrespective of the process occupying the pipeline.

\par Importantly, as either approach can be influenced through instruction execution, they can both be activated and de-activated at any time upon attacker's will.
To exemplify it, consider the general purpose registers or the operands of an adder in Table~\ref{table:cpu_tables}(a).
Both components can serve as hosts of IRT-1 and IRT-2 respectively, as both can be influenced by the execution of Assembly instructions (\textit{e.g.,} hold values mandated by the attacker), as per our evaluation in Section~\ref{sec:eval_cs}.


\subsection{Selecting Signals for the Trigger Mechanism}
\label{sec:tt_methodologies}
    In practice, settling on a way to implement a trigger's circuitry is non-trivial. 
To date, two major approaches exist.
\par First, the majority of implementations that target microprocessors (\cite{DBLP:KAFP19, DBLP:YHDAS16, ACM:VGHGSPJR, ACM:AHTPSPGS}) utilize so-called \textit{rare signals} (\textit{e.g.,} signals with a strong value bias, low toggling rate probability) for the creation of trigger circuits. 
The conjecture is that low toggling rate signals decrease the probability of inadvertently triggering the trojan.
To find these rare triggers, there has been an abundance of research  (\cite{ACM:VGHGSPJR, ACM:AHTPSPGS, DBLP:YHDAS16, YSLWOM19, CJHYMPBS18}) that use signals selected by profiling various benchmarks. 
Inevitably, the selection of rare signals is biased by the benchmark. 
Worse yet, silicon proven scenarios consider an adversary that has a priori knowledge of the workload(s) to be served by the microprocessor~\cite{DBLP:YHDAS16}, so that the benchmark used is representative of the common case.
Still, in such scenarios, if the workload served by the CPU changes during its life time, then the rareness of the signals might shift too.

\par What makes things more complicated is the fact that the selected rare signals must be translated to higher level instructions that can excite them.
Unfortunately, there has been little work (if any) on the development of methods that can offer this signal-to-instruction backwards compatibility and therefore this relationship should be derived through experimentation (\textit{e.g,} running sets of instructions and mapping them to signals excited).
Last, but not least, the use of rare signals at the fabrication stage imposes the extra burden of reverse engineering the logic associated with each one of the signals used, in order to figure how to excite them.

\par An alternative approach~\cite{DBLP:KTCGJZ08, AMVKADK, ACM:SPMGFSTRD} involves the monitoring of general purpose hardware for the existence of specific \textit{triggering values}.
The rationale is that by increasing the number of bits in the trigger values~\cite{ACM:VGHGSPJR} or the number of sequences used, one can reduce the chances of accidental triggering (we include an experiment about IRTs' susceptibility to inadvertent triggering in Section~\ref{sec:mitigations}).
Additionally, this method can be used without any a priori knowledge by the attacker of the likely to be executed CPU workloads.
Another practicality of this method during a fabrication stage attack, is that the attacker only needs to reverse engineer a single host module, assuming there are no multiple modules contributing to the implementation of the trigger circuit. 
It makes it also straightforward for the attacker to excite the trojan, as long as the host module of the trigger circuit can be directly influenced by instructions of the executed handling process.

\par For our implementation of IRTs, we choose the solution of \textit{triggering values} as we consider it to be more controllable and less incommodious to implement at the fabrication stage.
Therefore a common requirement for both of the IRTs' host modules, is for them to provide multi-bit signals that can participate in the generation of the trigger signal, like those in Table~\ref{table:cpu_tables}(a).
For the reader's convenience, we summarize in Table~\ref{table:irt_host_modules} all of the IRT host module requirements.


\begin{table}[h]
\centering
\begin{tabular}{c|c c c}
    \toprule
    &\multicolumn{3}{c}{\bfseries \textsc{Host Module Requirements}}\\
    \textbf{IRT Solution}&\bfseries \shortstack{Context Switch\\Support}&\bfseries \shortstack{Instruction\\Influenced}&\bfseries \shortstack{Multi-bit\\Signals}\\
    \midrule
    IRT-1& \textcolor{ForestGreen}{\checkmark}& \textcolor{ForestGreen}{\checkmark}& \textcolor{ForestGreen}{\checkmark}\footnotemark[1]\\
    IRT-2& \textcolor{Red}{\xmark}& \textcolor{ForestGreen}{\checkmark}& \textcolor{ForestGreen}{\checkmark}\footnotemark[1]\\
    \bottomrule
\end{tabular}
\caption{Host module requirements for IRT based trojans.}
\label{table:irt_host_modules}
\end{table}
\footnotetext[1]{If triggering values are selected as the triggering approach.}
\vspace{-1em}

\subsection{Tackling Silicon Reality Constraints}
\label{sec:tt_silicon_sol}
    \par Irrespective of the choice of triggering methodology, the fabrication stage attack scenario imposes the same practical constraints on the attacker. 
Namely, an adversary must find a way to overcome the fact that:
\begin{enumerate}[label=\it{\roman*)}]
    \item The trigger and the payload host modules might be in separated areas of the layout (\textit{no spatial locality}), thereby requiring the time critical trigger signal to travel a significant distance before reaching the payload.
    This in turn can lead to race conditions between the trigger signal and CPU events targeted by the payload. 
    
    \item The \textit{area or routing congestion} around host modules of interest might be high, leaving limited space for the placement and routing of trigger gates and nets.
    
    \item The signals of interest might be part of the circuit's \textit{critical path}, therefore further use of them in the HT circuits can lead to timing violations in the design.
\end{enumerate}

With respect to the spatial locality challenge, the IRT-2 solution is immune, as the extra state machine logic of the trigger circuit keeps the payload in an ``always-ready" state during context switching events.
However, the IRT-1 solution is susceptible to it, as the trigger signal needs to travel to the payload each time the CPU context switches back to the execution of the handling software.
To solve this challenge, the IRT-1 generated trigger will need to be treated as a \textit{multi-cycle path}.
In digital IC design, a multi-cycle path refers to a data path between two registers that operates at a sample rate lower than that of the clock signal.
This means that multiple clock cycles are available for a valid value to show up at the end of such a path.
To offer this option to the trigger signal, attackers can leverage time expensive CPU events to mask the time needed for the arrival of the trigger signal responsible for the payload setup.
In Table~\ref{table:cpu_tables}(b) we provide a list of the most conventional multi-cycle CPU events.
We consider these events to be potentials for attackers to gain extra clock cycles for the trigger signal propagation.
As a proof of concept, our implementation uses the \textit{page table walking} event to offer additional clock cycles to the trigger signal of an IRT-1 based HT in Section~\ref{sec:eval_cs}.

\par As noted earlier, attackers also face challenges with optimal layout integration of their trojans. 
Struggling to achieve tight performance constraints, high speed digital designs often face congestion issues either in parts of their placement or their routing.
Moreover, nets targeted by attackers for the generation of the trigger might belong to the critical path of the design.
This means that any additional logic attached to them can have an adverse effect on the timings of the design.
To prevail over such challenges, implementation of IRTs can benefit from the affluence of microarchitecture modules on which attackers can have immediate influence on, as seen in Table~\ref{table:cpu_tables}(a).
In our prototype, we attach IRT-1 and IRT-2 on a set of general purpose registers and the operands of the integer adder.
These host modules cover all of the requirements outlined in Table~\ref{table:irt_host_modules}.

\section{Evaluation}
\label{sec:eval_main}

\par To demonstrate the fact that IR-based trojans can operate in the presence of diverse context switching events while also overcoming the aforementioned silicon challenges, we couple our trojan designs with a persistent payload and attack a modern RISC-V microarchitecture.
For our experiments, we choose the CVA6 microarchitecture~\cite{ZB19,CVA6Git} and introduce a payload that can undermine the integrity and availability of a CPU design.
Specifically, we design a payload that violates the separation that the operating system mandates between privileged and non-privileged areas of a CPU's memory.
To achieve that, we target the exception generation mechanism within the memory management unit (MMU).

Our payload suppresses the exception signal generated by faulty store memory accesses that try to alter addresses whose privilege rights do not concur with the privilege state of the processor.
To do so, we interfere with the User-mode bit (U-bit) of the page table entry (PTE) under access, and present a modified U-bit version to the exception handling module. 

\par We use this payload to support two attacks. 
First, we perform an integrity attack that modifies arbitrary kernel space addresses selected by the attacker.
These addresses might belong to different kernel modules running on the CPU and are responsible for a diverse set of software security mechanisms (\textit{i.e.,} access control policies~\cite{LinuxKSM}, packet filtering).
To demonstrate the generalization of this attack, we perform an experiment in a controlled setting and target a custom made Linux kernel module (LKM).
Our LKM allocates a certain set of addresses inside the kernel address space and the attacker uses the handling process to modify the contents of the allocated addresses with attacker-specific values.  
We implement this attack with both IRT-1 and IRT-2 trojans.

\par Second, we perform an attack that affects availability.
In this experiment, we explicitly modify addresses containing kernel structures of type ``\textit{task\_struct}".
A ``\textit{task\_struct}" element is a process descriptor containing information about a respective process and belongs to the kernel's task list~\cite{LinuxKernelDev}.
In this particular attack we overwrite the addresses following the ``\textit{init\_task}" structure of the \textit{init} task and cause a kernel panic.


\subsection{Resilience to Interrupts}
\label{sec:eval_cs}

\par In what follows, we evaluate our IRTs under both kernel context switching and multitaksing scenarios.
We run our experiments using a CVA6 design with integrated IRTs, that is implemented on a Genessys 2 FPGA board.

\paragraph*{Kernel Context Switching}
Abiding by the requirements of Table~\ref{table:irt_host_modules}, we implement IRT-1 inside CVA6's register file and attach it on the nets of two of the core's 64-bit general purpose registers (GPRs).
The HT is enabled through a specific 128-bit sequence loaded in this set of registers.
To deactivate the HT completely, the sequence in the set of registers needs to be overwritten at the end of the attack.
In the mean time, a context switch out of the handling process means that the sequence will be temporarily removed from the register set and then restored once the handling software resumes execution.

\par In this experiment we measure the presence of kernel context switching during HT attacks.
Due to the fact that the CVA6 core is single threaded, a change of CPU privileges occurring during the execution of the HT handling process signifies a kernel context switch.
Moreover, to avoid mode switching (different than a full context switch), the handling software does not include any system calls.
Therefore, in this experiment, the CPU's transition to a state other than User-mode during the execution of the handling process can only happen upon a full kernel context switch. 
Table~\ref{table:Kernel_CS} shows the privilege level activity for differing amounts of data overwritten inside the kernel.
Although context switching between the kernel and the handling software happens throughout the execution of the attack, the IRT-1 design automatically re-enables the trigger signal every time the handling process resumes execution, as illustrated in Figure~\ref{fig:scheduling} and explained in the paragraph above.

\begin{table}[h]
\centering
\begin{tabular}{c c c c}
    \toprule
    \textbf{KBs}& \textbf{User Mode}& \textbf{Supervisor Mode}& \textbf{Machine Mode}\\
    \midrule
    0.5& 1& 0& 0\\
    1&  3& 7& 7\\
    4&  3& 7& 7\\
    16&  3& 7& 7\\
    32&  3& 9& 9\\
    \bottomrule
\end{tabular}
\caption{CPU privileges during kernel context switching.}
\label{table:Kernel_CS}
\end{table}

\begin{table}[h]
\centering
\begin{tabular}{c c c c}
    \toprule
    \textbf{KBs}& \textbf{User Mode}\footnotemark[2]& \textbf{Supervisor Mode}& \textbf{Machine Mode}\\
    \midrule
    0.5& 6& 20& 17\\
    1&  32& 97& 86\\
    4&  39& 116& 103\\
    16&  45& 135& 120\\
    32&  64& 192& 172\\
    \bottomrule
\end{tabular}
\caption{CPU privileges under multitasking conditions.}
\label{table:Multi_CS}
\end{table}
\footnotetext[2]{This number considers all user level process interleaved in the benchmark.}

\par This inevitable transition of the trigger signal from an ON state to an OFF state and vice versa, might seem to be dangerous to generate race conditions between the delivery of the payload and the CPU event that is targeted each time (\textit{e.g.,} in this case a faulty store memory access). 
This is a big challenge, especially in situations where the trigger circuit and the payload are placed in separated layout regions.
In such cases, the trigger signal will need to travel a long distance to reach the payload.
It should therefore be treated as a multi-cycle path, as discussed in Section~\ref{sec:tt_silicon_sol}.

\par In this attack scenario, to offer the extra clock cycles for the trigger to propagate to the payload, we take advantage of the page table walking event seen in Table~\ref{table:cpu_tables}(b).
As the targeted LKM addresses do not belong to the address space of the handling process, a page-table walk is invoked for each overwrite, in order to perform the virtual address to physical address translation.
Address translation always precedes the CPU's attempt to access the requested memory location.
A page-table walk requires multiple clock cycles to finish and consequently provides sufficient time for the trigger signal to propagate and enable the payload in the MMU.
Therefore, in a scenario where the CPU has resumed the execution of the handling software, the payload is setup (re-enabled) while the page table walking is in effect.
Then once a faulty store memory access inside the kernel address space occurs, the payload is able to suppress the generated exception signal of the MMU.
This way the handling software manages to successfully overwrite the requested kernel space addresses.

\paragraph*{Process Context Switching}
For the implementation of IRT-2 we go after the requirements of Table~\ref{table:irt_host_modules} and use the integer adder in the arithmetic logic unit (ALU) of CVA6 as the host module.
We attach IRT-2 on the nets of the two 64-bit operand inputs of the integer adder.
At the start of the attack, we influence the adder's operands with a specific set of \textit{activation} values, which forces the FSM of Figure~\ref{fig:always_readyFSM} to state S1 and enables the trigger.
The trigger remains as such until the end of the attack, where a specific set of \textit{de-activation} values is loaded at the operands and completely disables the trigger (FSM switches to S0).

\begin{figure}[ht]
\begin{subfigure}{.5\textwidth}
  \centering
  \includegraphics[width=0.9\linewidth]{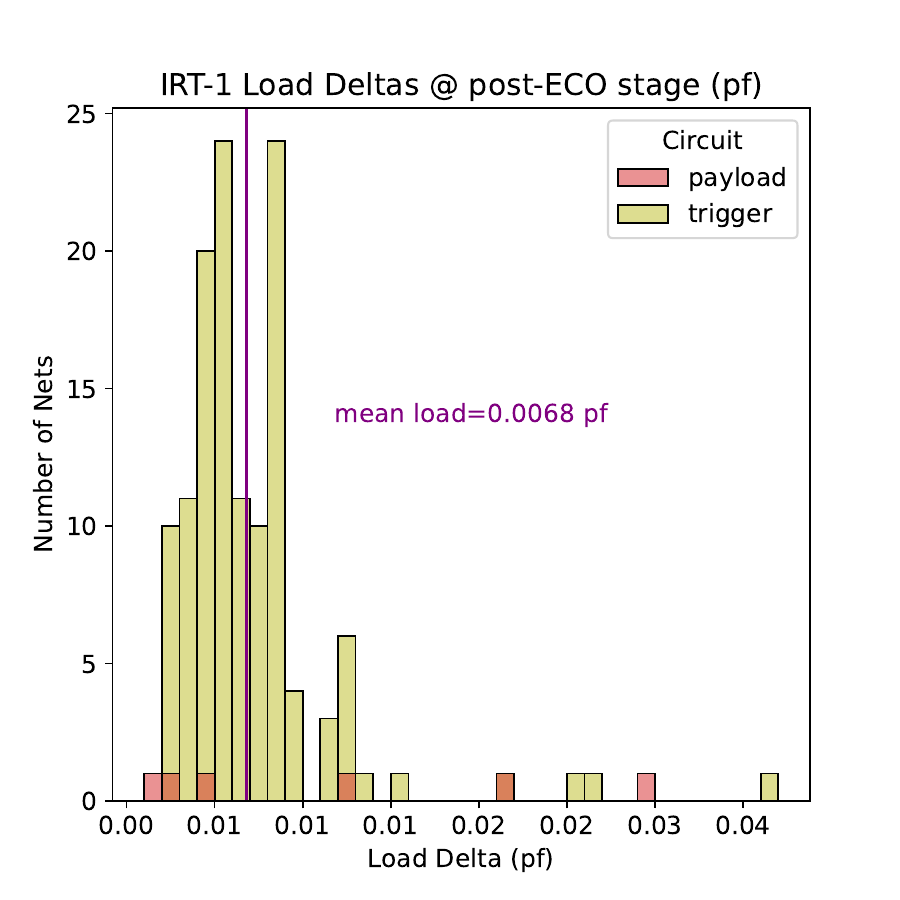}  
  \caption{IRT-1}
  \label{fig:IRT1load}
  \vspace{-2em}
\end{subfigure}
\begin{subfigure}{.5\textwidth}
  \centering
  \includegraphics[width=0.9\linewidth]{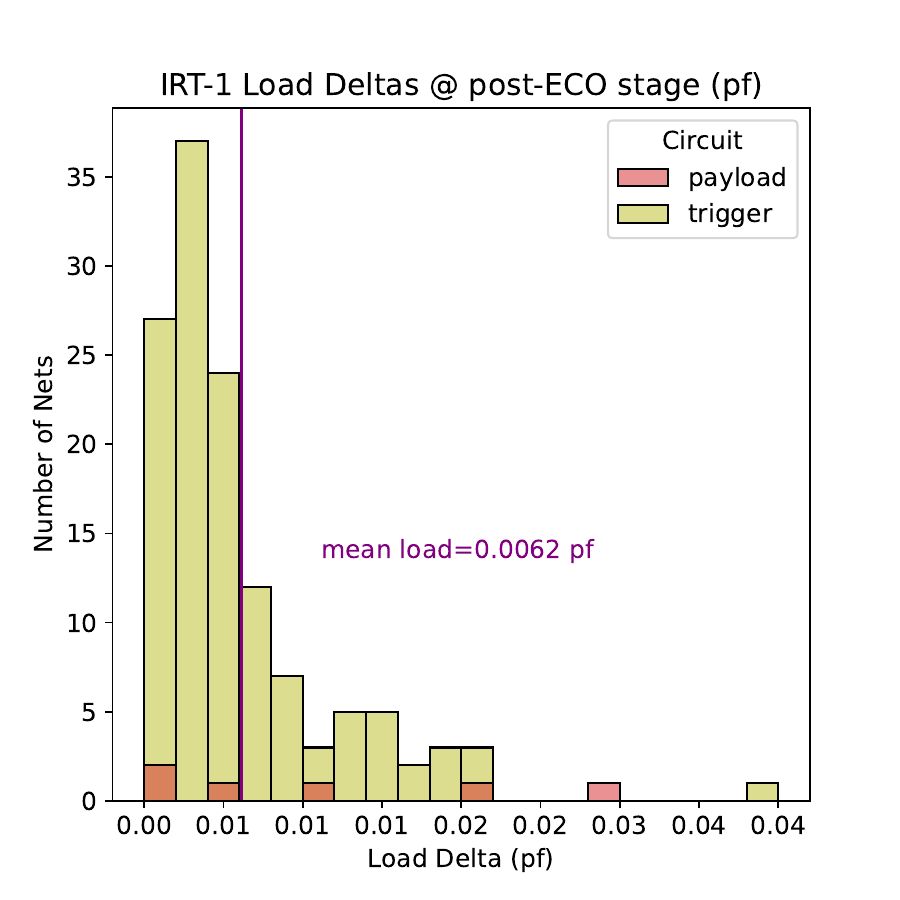}  
  \caption{IRT-2}
  \label{fig:IRT2load}
\end{subfigure}
\caption{Load (pf) overhead on original layout nets.}
\label{fig:IRTload}
\vspace{-1em}
\end{figure}
\begin{figure}[ht]
\begin{subfigure}{.5\textwidth}
  \centering
  \includegraphics[width=0.9\linewidth]{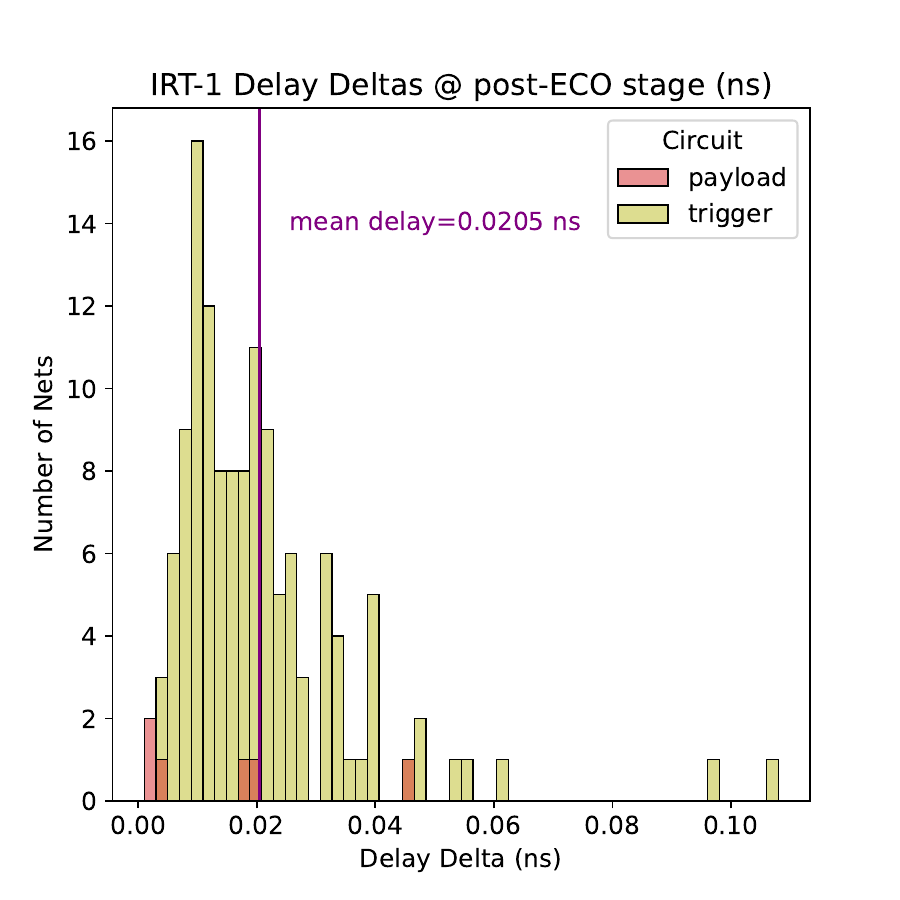}  
  \caption{IRT-1}
  \label{fig:IRT1delay}
  \vspace{-2em}
\end{subfigure}
\begin{subfigure}{.5\textwidth}
  \centering
  \includegraphics[width=0.9\linewidth]{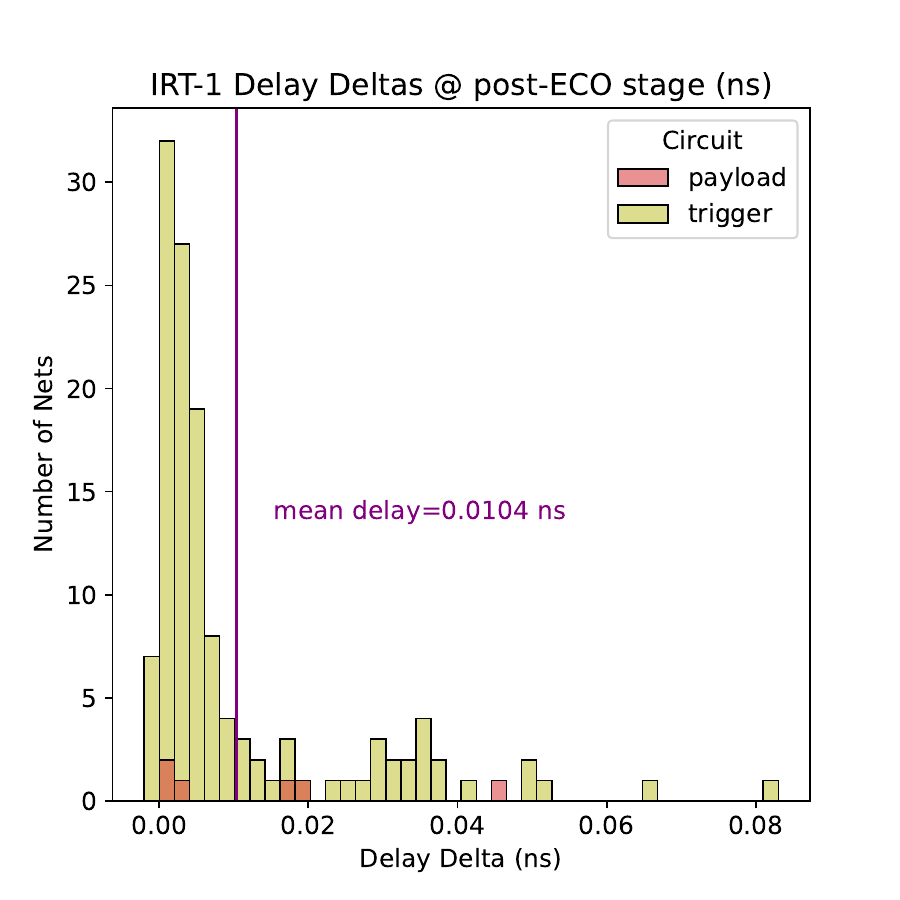}  
  \caption{IRT-2}
  \label{fig:IRT2delay}
\end{subfigure}
\caption{Delay (ns) overhead on original layout nets.}
\label{fig:IRTdelay}
\vspace{-1em}
\end{figure}

\par In this experiment we measure the prevalence of interrupts due to multitasking during the execution of a HT attack.
Since CVA6 is a single thread core, context switching due to multitasking cannot be tested directly.
However, we emulate the multitasking operation by having our HT handling process being interleaved with the execution of other user level processes and the kernel.
First, we set the IRT-2 generated trigger to an `always-ready' state, as explained above and then we execute a program that randomly interleaves the execution of the handling process (that overwrites arbitrary addresses in the kernel module) with the execution of a general computing performance benchmark (called Dhrystone).
Table~\ref{table:Multi_CS} shows the privilege level activity for different number of kilobytes overwritten inside the kernel by the HT handling process.
Despite the heavy kernel and process context switching happening in the pipeline, the payload supresses the generation of exceptions upon faulty store memory accesses that happen whenever the CPU context switches to the handling software.
This is due to the generated trigger now being active irrespective of the process executed in the pipeline, until the attack is complete.

Taken together, these experiments aptly show that context switching is a prevalent phenomenon that can impact the execution of HT attacks.
Therefore, practical attacks must account for these real-world behaviors in their designs.
Our experiments indicate that IRT designs can be a reliable solution that attackers can use to prevail over such phenomena.

\subsection{Reassessing Fabrication Stage Flexibility}
\label{sec:eval_asic}
    \par
\begin{figure}[ht]
    \centering
    \includegraphics[scale=0.028]{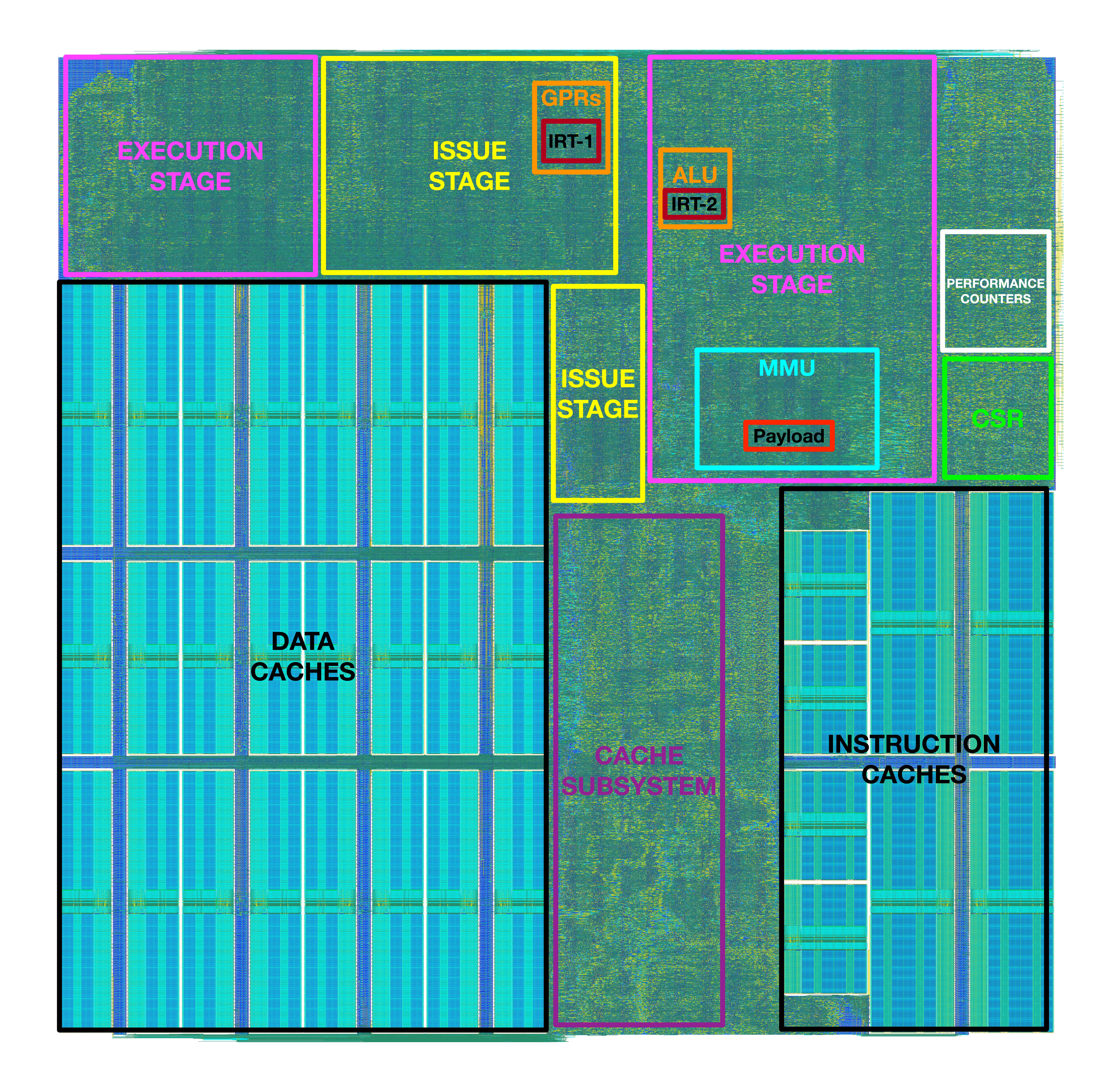}
    \caption{\centering Tape-out ready layout depicting both HT designs.}
    \label{fig:cva6_modules}
    \vspace{-1em}  
\end{figure}

 To provide supporting evidence for our assertion that IRTs provide a pathway for fabrication stage attacks, we use the CVA6 tape-out ready layout (sign-off timings and no manufacturing violations) shown in Figure~\ref{fig:cva6_modules} to insert our trojans and measure their impact on the host modules and the overall layout performance.
The CVA6 layout of Figure~\ref{fig:cva6_modules} is both high speed (clock frequency of 600MHz) and high density ($>$75\% core utilization), making it demanding to place and route the extra attack gates and nets.
The layout is implemented in a commercial 28nm technology process and is free of trojans, representing the GDSII file received at the foundry from a trusted design house. 

\begin{table*}[ht]
\footnotesize
\centering
\scalebox{0.8}{
\begin{tabular}{c|c c c c c c c| c c c c c}
    \toprule
    &\multicolumn{5}{c}{\bfseries \textsc{Hardware Trojan Characteristics}} &\multicolumn{5}{c}{\bfseries \textsc{Physical Implementation Results}}\\
    \bfseries Target&\bfseries Trigger&\bfseries \shortstack{\# Trigger\\Bits}&\bfseries \shortstack{Trigger\\Host}&\bfseries \shortstack{Payload\\Host}&\bfseries \shortstack{\# Comb.\\Cells}&\bfseries \shortstack{\# Seq.\\Cells}&\bfseries \shortstack{\# Conn.\\Nets}\footnotemark[3]&\bfseries \shortstack{Frequency (MHz)\\before}&\bfseries \shortstack{Density (\%)\\before \text{\textbar} after}&\bfseries \shortstack{Total Power ($\mu$W)\\before \text{\textbar} after}& \bfseries \shortstack{Slack (ps)\\before \text{\textbar} after}& \bfseries \shortstack{\# Violations\\after}\\
    \midrule
    CVA6& IRT-1& 128&Set of GPRs& MMU& 50& 4& 55& 600& 75.32 \text{\textbar} 76.537& 241.31 \text{\textbar} 241.44& 0.00 \text{\textbar} 0.00& 0\\
         & IRT-2& 128&ALU Adder& MMU& 64& 6& 70& 600& 75.32 \text{\textbar} 76.539& 241.31 \text{\textbar} 244.70& 0.00 \text{\textbar} 0.00& 0\\
    \bottomrule
\end{tabular}
}
\caption{Physical implementation results showing the impact of IR-based HTs on a CVA6 layout after insertion.}
\label{table:cva6_elem}
\end{table*}

\par For the design of the trigger circuits, we follow the strategy suggested by~\citet{SSLY21} wherein they analyze different logic gate connection patterns and record the ones that exhibit very low transition probability on their output signal.
Such gate connection patterns can produce trigger circuits with very low (inadvertent) activation probability, therefore enhancing the HTs' overall stealthiness (\textit{e.g.,} a dormant HT does not affect the CPU's operation and therefore is harder to detect).
We use the low transition probability connection patterns $\langle {\tt AND} \rightarrow {\tt NAND} \rangle, \langle {\tt NAND} \rightarrow {\tt NOR} \rangle$
as the core of IRT-1 and IRT-2 trigger circuits.
For more information about the low probability connection patterns, we refer the reader to~\cite{SSLY21}.
We use two separate CVA6 layout versions to insert the IRT-1 and IRT-2 triggers combined with the payload of Section~\ref{sec:eval_main}.
\footnotetext[3]{This is the number of the interconnects between the HT gates.}

\par Our threat model considers an adversary that gains access to the finalized layout at the foundry to perform an IRT insertion.
The IRT is inserted in the open areas of the layout near the modules of interest (\textit{e.g.,} MMU, GPRs, ALU).
To emulate \textit{the experience and expertise of a skilled adversary attempting a manual IRT insertion} we use the pre-mask ECO flow that P\&R tools provide.
The reasons for this choice are twofold:
\begin{enumerate}[label=\it{\roman*)}]
    \item ECOs keep the existing layout intact and remove only filler cells for the placement of the attack gates.
    \item ECOs avoid, to the extent possible, deterioration of the layout's timings during routing of attack nets.
\end{enumerate}
An important concern for the adversary is the coupling capacitance inserted from the HT wiring.
The use of ECOs clearly enables minimizing the impact of it.
In an actual manual insertion, we argue that the small number of gates and the low number of routing wires for IRTs (Table~\ref{table:cva6_elem}) work in favor of the attackers in their attempt to keep the inserted coupling capacitance from HT wiring to minimal levels.
Besides, coupling capacitance is a well studied topic in the microelectronics community~\cite{EMABMA03}, with reduction techniques that an attacker can utilize (e.g., wire sizing, wire spacing, adding repeaters and jogging to different metal layers).
For the reasons mentioned, \textit{we consider our use of ECOs to resemble a realistic manual HT insertion from a knowledgeable adversary}.
After the HT insertion, we verify that the CVA6 layout continues to meet the initial fabrication standards, through sign-off static timing analysis and DRC checks. 

Table~\ref{table:cva6_elem} summarizes the impact IR-based HTs have on the CVA6 layout in terms of density, power consumption, critical path slack and design rule violations.
The discrepancy seen in the number of attack cells between the two IRTs is attributed to the extra support logic required for IRT-2 (Figure~\ref{fig:always_readyFSM}) and the repeater cells required for the trigger signal of the further away placed IRT-1 (Figure~\ref{fig:cva6_modules}).
Our results are comparable to those presented by~\citet{ACM:AHTPSPGS}. 
In particular, although CVA6's layout has a more strict critical path (0ns versus 15ns slack for PULPino in~\cite{ACM:AHTPSPGS}), IRTs do not introduce any timing violations (unlike PULPino's case).

\par To further assess the feasibility of fabrication stage IRTs, we compute the capacitive load overhead (see Figures~\ref{fig:IRT1load} and~\ref{fig:IRT2load}) and the corresponding time delay overhead (see Figures~\ref{fig:IRT1delay} and~\ref{fig:IRT2delay}) on the nets that IRTs' are attached on (GPR's and adder operands in Section~\ref{sec:eval_cs}).
The average load added by IRT-1's design on this layout and technology process is 6.8fF (6.2fF for IRT-2) which yields an average extra delay on the nets of 20ps (10ps for IRT-2).
The very low overhead suggests that IRTs can target very high frequency CPU layouts (\textit{e.g.,} IRT-1's 20ps overhead is only 3.4\% of the original 1.7GHz clock cycle in \cite{ZB19}).
Overall, our measurements show that IRT designs can be minimalistic enough for fabrication attacks.


\paragraph{Microarchitecture Discussion}
Lastly, we note that \citet{ZB19} designed the CVA6 microarchitecture with a register file (\textit{e.g.,} integer and floating point registers) that is based on flip-flops.
That microarchitectural choice allows us to implement IRT-1 inside the CVA6 register file, as we can use the nets coming out of the flip-flops for the generation of the trojan trigger.
A different practice featured in state-of-the-art microarchitectures is to use SRAM-based register files obtained through commercial memory compilers.
The structure of IRT-1 cannot be implemented in such SRAM-based register files, but that limitation does not detract from the generality of IRT attacks, in the same way that the requirement for a single-bit register to escalate the privilege rights of a user process in the OR1200 microarchitecture~\cite{OR1200Git} does not detract from the generality of the A2 trojan attacks by~\citet{DBLP:YHDAS16}.
Besides, for microarchitectures with SRAM-based register files, attackers can seek different pathways to implement IRTs by targeting a different general purpose module in Table~\ref{table:cpu_tables}(a).

\section{Mitigations}
\label{sec:mitigations}



\par Comprehensive inspection of fabricated dies at large scale is considered to be extremely difficult and expensive.
Adding to that, post-fabrication detection of hardware trojans is challenging because fabricated chips are like black boxes in the eyes of an evaluator. 
Therefore evaluators need to come up with clever strategies in order to detect trojans at scale. 

\par One defense might be to use test-pattern generation techniques that excite rare logic conditions of the design under test.
To that extent, one could utilize existing methods~\cite{CRSWFPSPCBW09,ACM:LYMP21MAX,IEEE:LYMP21TAR,ACM:HYBSMP16} to classify the rareness of signals in digital designs, and subsequently use the classification to generate high coverage test patterns (\textit{e.g.,} the number of rare nets excited).
The goal of the specially-crafted patterns is to either produce an out-of-spec behavior of the IC or increase the side-channel information leaked by the introduced HTs.
However, as noted earlier, our IRT trigger mechanisms utilize nets of general purpose hardware that are frequently excited.
Because such nets cannot be considered rare \cite{ACM:VGHGSPJR}, existing methods can not be readily applied.

\begin{figure}[ht]
    \centering
    \includegraphics[width=1\columnwidth]{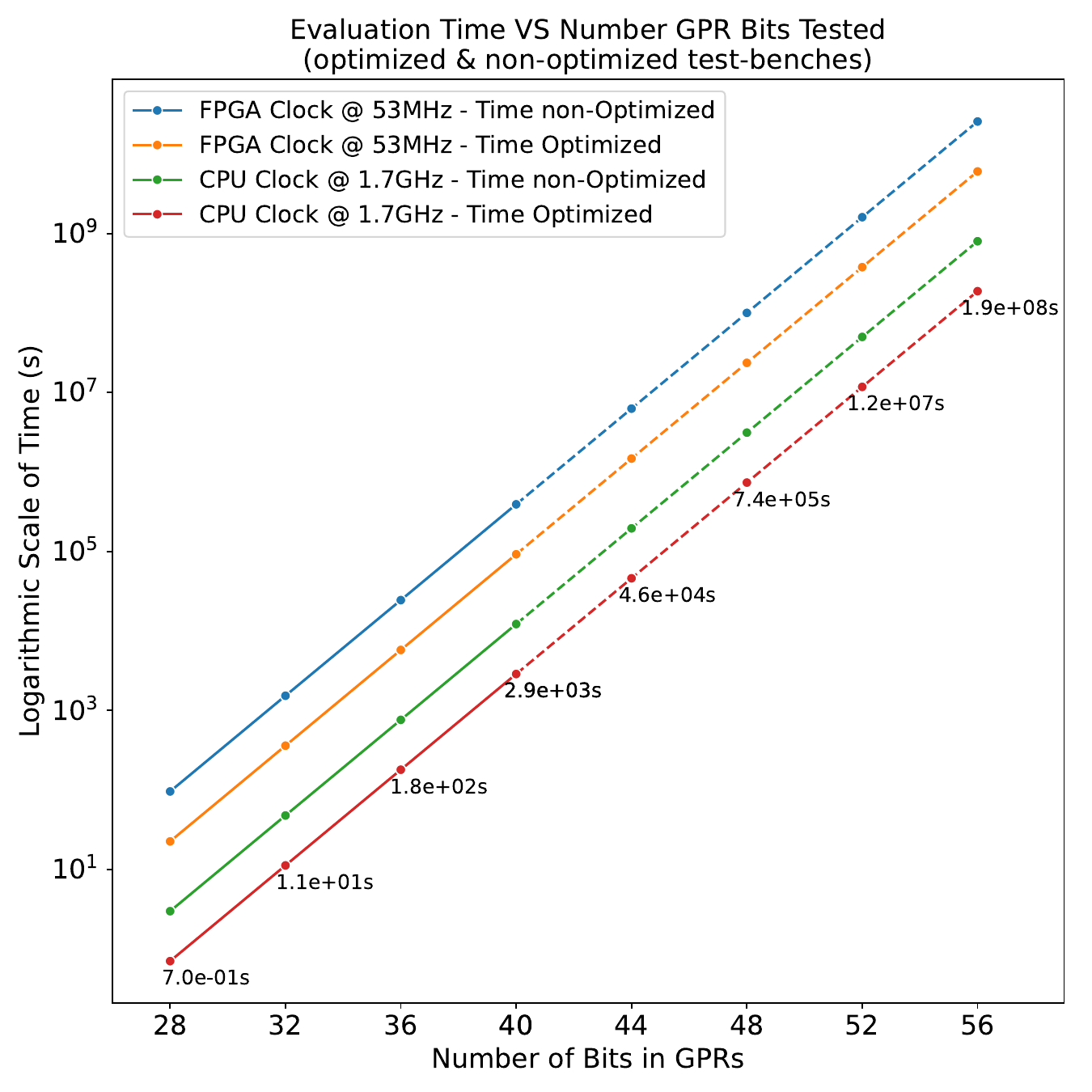}
    \caption{\centering Runtime for different number of excited GPR bits.}
    \label{fig:fpga_bits_experiment}
      \vspace{-1em}
\end{figure}

\par In the functional testing domain, a brute force method might seem like a viable strategy when looking to excite HTs with trigger circuits based on triggering values.
But, such a strategy is likely to be impractical. 
To demonstrate that, we use the FPGA implementation of CVA6 and run on Linux a test-bench sequentially incrementing two general purpose registers by 1, for different bit widths.
We view this as one of the simplest test-bench strategies an evaluator might run to excite an IRT-1 flavored trojan.
Two versions of this experiment were conducted (with and without GCC compiler optimizations) for widths up to 40-bits (denoted by solid lines in Figure~\ref{fig:fpga_bits_experiment}).
The measurements exhibit a $16\times$ linear factor on a logarithmic time scale.
We use this factor to infer the equivalent time for values with sizes greater than 40-bits (shown as dashed lines).
Since the FPGA's clock frequency is at 53MHz, we extrapolate the measurement's time to the CVA6's native frequency of 1.7GHz.
The result show that, in practice, this is not a viable defense: looping just over $2^{48}$ values takes roughly 9 days.

\par In the long run, detection techniques that leverage side-channel~\cite{DBLP:NguyenCPZ19} methods or scanning electron microscope (SEM) methods~\cite{EPTMSBCKAMCP23, VNLHSQRMTSHWDLANTM18, CFLMPFJJATA15} offer promising directions for detecting trojans inserted during fabrication. 
SEM methods are particularly powerful, as they are based on the complete de-layering of a chip.
As metal layers are removed, high precision images of the layout are captured and then compared with the respective parts of the original GDSII file sent for fabrication.
This meticulous procedure is very expensive in terms of equipment and time, and the tested chips end up completely destroyed.
Therefore, SEM techniques can be used only on a small percentage of the fabricated chips.
Nevertheless, SEM techniques provide strong guarantees of discovering fabrication attacks based on swapping of filler cells with HT gates~\cite{EPTMSBCKAMCP23}.

\par With that in mind, one way to \textit{indirectly} maximize the coverage of SEM methods, in order to test more chips, is potentially through the use of side-channel detection methods as the one by \citet{DBLP:NguyenCPZ19}.
The major drawback of such side-channel methods is the need for reference measurements captured from a golden IC (e.g., an IC generated through a fully trusted design and fabrication cycle).
SEM methods can prove valuable here, as an evaluator can use side-channel methods to get the EM signature of a chip and then inspect the chip with SEM to prove it is tamper-proof.
If that is the case, then the collected measurement can serve as the golden reference for comparison with the measurements collected from the rest of the fabricated chips.

\section{Conclusions}
\label{sec:conclusions}

In this work we measure the prevalence of context switching events that happen during the CPU's normal operation and can render unreliable the successful execution of HT attacks.
We therefore argue that attack reliability is synonymous to attack stealthiness.
Motivated by this observation, we introduce a new class of CPU-oriented trojan trigger circuits, called IRTs. 
Resilient by nature against the non-deterministic context switching events measured, IRTs can reliably deliver purposeful payloads against their host CPU.
Furthermore, our work raises questions around the perception of flexibility for trojan insertion at the fabrication stage, as we show that IRT designs can be integrated efficiently inside tape-out ready layouts.  
Since suitable IRT hosts are modules estranged from the implementation of hardware enforced security policies, we reinforce the observations of~\citet{KDJC23} about the need of securing general purpose hardware in CPU designs.
To promote further work in this area, we open-source our designs and the supporting control software in~\cite{IRTGit}.

\bibliographystyle{abbrvnat}
\bibliography{references}

\end{document}